\documentclass[sigconf, screen=true]{acmart}

\usepackage[yyyymmdd]{datetime}

\usepackage{tabularx}
\usepackage{dcolumn} 
\newcolumntype{d}[1]{D{.}{.}{#1}}

\usepackage{subcaption}

\usepackage{todonotes}
\let\xtodo\todo
\renewcommand{\todo}[1]{\xtodo[inline,color=green!50]{#1}}

\AtBeginDocument{%
  }

\setcopyright{cc}
\copyrightyear{2025}
\acmYear{2025}
\acmDOI{10.48550/arXiv.2504.04253}

\acmConference[]{Workshop on AI Tools for Thought at CHI 2025}{April 26}{Yokohama, Japan}

\begin{document}

\title{User-Centered AI for Data Exploration: Rethinking GenAI’s Role in Visualization}


\settopmatter{authorsperrow=2}

\author{Kathrin Schnizer} 
\orcid{0009-0007-6952-2340}
\affiliation{
  \institution{LMU Munich}
  \city{Munich}
  \postcode{80337}
  \country{Germany}}
\email{kathrin.schnizer@ifi.lmu.de}

\author{Sven Mayer}
\orcid{0000-0001-5462-8782}
\affiliation{%
  \institution{LMU Munich}
  \city{Munich}
  \postcode{80337}
  \country{Germany}}
\email{info@sven-mayer.com}


\begin{abstract}
Recent advances in GenAI have enabled automation in data visualization, allowing users to generate visual representations using natural language. However, existing systems primarily focus on automation, overlooking users' varying expertise levels and analytical needs. In this position paper, we advocate for a shift toward adaptive GenAI-driven visualization tools that tailor interactions, reasoning, and visualizations to individual users. We first review existing automation-focused approaches and highlight their limitations. We then introduce methods for assessing user expertise, as well as key open challenges and research questions that must be addressed to allow for an adaptive approach. Finally, we present our vision for a user-centered system that leverages GenAI not only for automation but as an intelligent collaborator in visual data exploration. Our perspective contributes to the broader discussion on designing GenAI-based systems that enhance human cognition by dynamically adapting to the user, ultimately advancing toward systems that promote augmented cognition.
\end{abstract}




\begin{teaserfigure}
    \centering
    \makebox[\textwidth][c]{\includegraphics[width=1\textwidth]{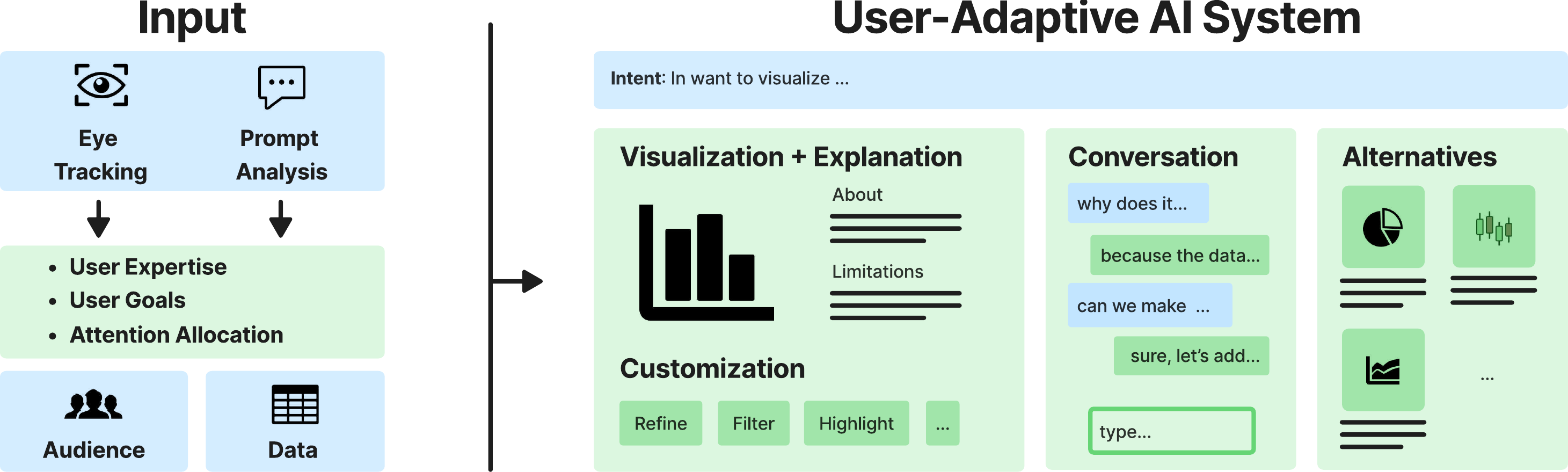}}
    \caption{Schematic overview of the proposed user-adaptive AI system for data exploration and visualization. The system leverages eye tracking and prompt analysis to model user expertise, goals, and attention allocation, enabling personalized responses. It further considers the target audience and the underlying data to inform both interaction and visualization design. By offering transparent explanations, customization options, conversational refinement, and alternative suggestions, the system promotes a collaborative, iterative data exploration process that encourages curiosity and critical thinking.}
    \label{fig:vision}
\end{teaserfigure}

\maketitle

\vspace{-3mm}
\section{Introduction}
The rapid adoption of Generative AI (GenAI) is reshaping workflows across domains, which is driven by advancements in natural language processing and the ability to interpret and generate visual content. Initially, GenAI’s breakthroughs in text generation transformed how humans interact with information; now, its expanding capability to generate images, videos, and other visual media is redefining how we engage with information. As these systems become increasingly proficient at automating complex cognitive tasks, concerns arise regarding their impact on human cognition—specifically, as they might reduce the need for critical thinking and decision-making. By taking over processes once performed manually, GenAI imposes the risk of shifting humans from active sensemakers to passive consumers of AI-generated insights~\cite{lee2025impact}.

One domain where this tension between AI-driven automation and human cognition is particularly prominent is data visualization, an area explicitly designed to support sensemaking and critical reasoning. As data volumes continue to grow, our ability to interpret and extract meaning from complex information is crucial for decision-making in science, business, and public discourse. Data visualization serves as a powerful cognitive augmentation tool, enabling individuals to detect patterns, generate insights, and communicate findings more effectively than raw data alone.   
This makes data visualization a particularly compelling case study for examining how GenAI can augment cognition. Visualizations are inherently designed to enhance human reasoning, raising a key question: How can AI further strengthen cognition in a domain that is already a cognitive augmentation tool for recognizing patterns and generating insights? If designed without care, GenAI risks shifting visualization from an active reasoning process to a passive experience, where insights are merely consumed rather than actively constructed and critically examined. We believe that addressing this challenge is essential to understanding the broader implications of GenAI on human cognitive engagement.  

While recent efforts have integrated GenAI into visualization workflows, e.g., automating the creation of charts~\cite{dibia2023lida, wang2023llm4vis, tian2024chartgpt, narechania2020nl4dv, liu2021advisor, maddigan2023chat2vis} and sometimes combined with natural language descriptions~\cite{dibia2023lida, wang2023llm4vis}, these developments have largely prioritized efficiency over cognitive engagement. Current AI-driven visualization tools operate under the assumption that chart generation is synonymous with understanding it, overlooking the cognitive steps of forming an analytical intent and critically interpreting the visualized data. However, data visualization is not solely a means of displaying information. It is a process of sensemaking that involves the ability to critically interpret, interact with, and derive meaning from visualizations~\cite{boy2014Principled}.

This position paper advocates for an adaptive approach in GenAI-based visualization tools, where we tailor interactions, reasoning, and outputs to the user’s level of expertise.  We begin by reviewing existing automation-driven tools. Next, we outline potential methods for assessing user expertise, examine their limitations, and highlight key open questions that call for interdisciplinary discussion. Our goal is to contribute an example of how to take a user-centric perspective in this dialog, emphasizing the importance of incorporating the user into the design process of GenAI-based systems. To this end, we present a holistic vision for a system that augments cognition, moving beyond pure automation to leverage GenAI as a critical collaborator in visual data exploration.

\section{Current Data-First Automation Systems}
Next, we review existing GenAI systems, highlighting their capabilities and limitations to motivate a shift toward a more user-centered perspective. One that actively considers the user in the interaction to achieve meaningful and tailored cognitive augmentation.

\subsection{Overview of Recent Systems}
Existing GenAI-based visualization systems leverage different types of inputs. Most systems take tabular data as their primary input~\cite{dibia2023lida, wang2023llm4vis, liu2021advisor, tian2024chartgpt, maddigan2023chat2vis, narechania2020nl4dv, song2022rgvisnet, luo2021natural, vaithilingam2024dynavis, wisiecka2022comparison}. Beyond raw data, some approaches allow users to specify high-level visualization goals in natural language, focusing on the insights they want to gain rather than defining precise visualization specifications~\cite{tian2024chartgpt, liu2021advisor, narechania2020nl4dv, maddigan2023chat2vis}. Other systems focus on natural language queries for modifying existing visualizations, enabling users to refine visual encodings, add annotations, or adjust filters through iterative interactions~\cite{luo2021natural, vaithilingam2024dynavis, wisiecka2022comparison}.

The outputs generated by these systems vary in complexity and format. Many approaches generate code to render visualizations~\cite{dibia2023lida, wang2023llm4vis, liu2021advisor, luo2021natural, maddigan2023chat2vis, narechania2020nl4dv, vaithilingam2024dynavis, wisiecka2022comparison}. Additionally, certain systems enhance transparency by providing reasoning behind the generated visualizations or textual summaries of the data, explaining why a particular visualization was chosen or summarizing key insights~\cite{dibia2023lida, wang2023llm4vis}.

To assess the performance of these systems, various technical evaluations have been conducted, often focusing on accuracy~\cite{wang2023llm4vis, luo2021natural, liu2021advisor, song2022rgvisnet}, which examines whether the input data is correctly translated into the visualization, and reliability~\cite{dibia2023lida, tian2024chartgpt, maddigan2023chat2vis, song2022rgvisnet}, which evaluates the consistency of outputs across repeated queries. Only few systems have been evaluated through user studies to examine usability and interaction effectiveness~\cite{luo2021natural, wisiecka2022comparison, vaithilingam2024dynavis}.

Despite advancements in integrating GenAI into the workflow of generating data visualizations, several limitations remain. First, some approaches still lack standardized human evaluation, making it difficult to assess their usability and effectiveness in real-world applications~\cite{dibia2023lida, wang2023llm4vis, liu2021advisor, narechania2020nl4dv}. Additionally, varying accuracy in natural language interpretation was reported, as the systems were struggling to correctly map user queries to the appropriate commands for visualization generation~\cite{tian2024chartgpt, maddigan2023chat2vis, narechania2020nl4dv, song2022rgvisnet}. Lastly, some approaches do not support follow-up queries, limiting users' ability to iteratively refine visualizations or explore alternative designs~\cite{song2022rgvisnet, narechania2020nl4dv}.

\subsection{Remaining Gaps}
Concluding the overview of existing systems that incorporate GenAI into the generation of data visualization, they all address the key opportunity that natural language input allows users to bypass the need to translate their intent into specific visualization grammars, making the process more intuitive and accessible. Furthermore, natural language-based systems can accommodate the diverse ways individuals express their intentions, enabling more flexible interactions~\cite{wisiecka2022comparison, narechania2020nl4dv, tian2024chartgpt}. However, current GenAI-driven visualization tools remain mostly automation-focused, generating technically valid charts while often overlooking the expertise, needs, and high-level intentions of the human interacting with them. While some approaches allow users to specify their analytical intent~\cite{tian2024chartgpt, liu2021advisor, maddigan2023chat2vis, narechania2020nl4dv}, they generally lack a broader human-centered perspective. These systems typically function in a one-directional manner—producing visualizations without striving to promote deeper user engagement or critical reflection. Thus, a key open challenge is how to move beyond automation-driven visualization generation toward systems that actively account for the individual user and their unique needs.

\section{Towards Expertise-Aware Adaptive Systems} 
How can we move GenAI visualization tools beyond automation to truly augment human cognition? How can we empower users as active sensemakers rather than passive GenAI output consumers? We argue that visualization systems must be human-centered. By embedding awareness of the user and their interaction with the system, they could dynamically adapt to expertise levels while communicating their capabilities and limitations.

To effectively support cognitive augmentation, the system must account for the user's expertise in data visualization and analysis. Research has shown that adapting task difficulty enhances engagement and learning~\cite{klinkenberg2011computer} and that explanation effectiveness varies with user proficiency~\cite{szymanski2021visual}. Building on these insights, we explore how adaptive GenAI-based visualization systems could assess and integrate user expertise into their functionality. We cover the opportunities and challenges of the suggested approaches and conclude by highlighting the remaining open questions. 

\subsection{Assessment and Adaptation to Expertise} 
Inferring a user’s level of expertise can be approached in multiple ways, each with its own limitations. Self-assessment of users' data literacy is often prone to inaccuracies~\cite{kim2016people}. Alternatively, users could complete a visualization literacy assessment such as the VLAT~\cite{lee2016vlat} or Mini-VLAT~\cite{pandey2023mini}, which results in a proficiency score. However, these assessments are time-consuming, disrupt workflow, and may still be unreliable due to guessing the correct answer.  

A dynamic approach lies in inferring expertise from user behavior. The use of natural language creates the opportunity to analyze the prompts, examining conciseness, ambiguity, and complexity to distinguish between vague exploratory questions and precise analytical requests. A conversational system could further refine its assessment by prompting users for clarification where needed. Another strategy is to observe user interactions, identifying patterns such as frequent rejection of AI-generated visualizations followed by manual refinements, which could indicate expertise, or a strong reliance on system-generated suggestions, which may suggest a novice user. More advanced techniques, such as measuring cognitive workload through pupil dilation~\cite{pomplun2019pupil}, could further contribute to user assessment. However, this method requires specialized hardware for reliable results, as webcams exhibit higher measurement errors~\cite{wisiecka2022comparison}. Additionally, pupil dilation is sensitive to changes in illumination~\cite{kramer2020physiological}, requiring stable lighting for reliable results.

Effectively adapting to user expertise according to the user's behavior presents the key challenge to risk misinterpreting user behavior. Frequent modifications to a visualization could signal either expertise or a lack of comprehension, making it difficult for the system to infer the appropriate level of support. For example, a user who repeatedly refines a chart may be fine-tuning it based on deep analytical reasoning, demonstrating advanced understanding. On the other hand, another user making similar modifications might be struggling to find a clear way to represent the data, indicating a need for additional guidance. Without deeper contextual awareness, the system risks making incorrect adjustments, such as restricting advanced features for an expert experimenting with design variations or overwhelming a novice who is already struggling with comprehension.  

\subsection{Open Questions}  
Despite the opportunities that adaptive systems create, several fundamental research questions remain open. For instance, how can we ensure that inferred expertise measures reliably reflect real-world competencies? Developing assessment methods that accurately capture user proficiency without introducing bias or friction remains an open challenge.
Another key research question in adaptive system design for GenAI is how to balance personalized guidance with opportunities for learning and skill development. While novices may benefit from structured support, experts should be encouraged to deepen their analysis rather than be constrained by oversimplified explanations. However, if a system adapts too strongly to perceived expertise, it risks either limiting opportunities for learning or failing to provide adequate support. The challenge lies in developing an adaptive system that not only tailors guidance to a user’s skill level but also strategically encourages them to step beyond their comfort zone, fostering deeper engagement and new analytical perspectives.
Finally, another unresolved aspect involves considering not only the expertise of the visualization creator but also that of the intended audience. While systems can adapt to the user's level of knowledge, an effective visualization must also align with the proficiency of its consumers, which most likely will not be the same. Taking this full lifecycle of visualization into consideration, we are confronted with further questions, such as: Should visualizations be tailored to the least proficient viewer to ensure accessibility, or should they prioritize conveying deeper insights for more advanced audiences? 

\section{Our Vision}
Building on the outlined opportunities to improve GenAI-based approaches for cognitive augmentation, we take it a step further by presenting our vision for a human-aware system that dynamically adapts to user expertise, fostering deeper and more critical engagement with data. By shifting from a passive automation tool to an active collaborator, the system not only assists in visualization creation but also challenges users’ assumptions, encourages exploration, and promotes analytical thinking. To achieve this, the system could infer user expertise through prompt analysis and eye-tracking, allowing it to adapt its guidance and explanations accordingly, as visualized in \autoref{fig:vision}.

Rather than simply generating valid charts from a given dataset, the system should engage users in an iterative refinement process. Users can specify their intent using natural language, with the system offering structured guidance in refining their visualizations. Then, AI-generated suggestions are dynamically adjusted based on user input, and the interface makes reasoning explicit through transparent explanations. By continuously assessing user interactions, such as which elements they focus on, how they refine their queries, or where they hesitate, the system could adapt its feedback to provide the right level of detail at the right time. To ensure flexibility, users should be able to manipulate key aspects of the visualization, such as filtering data, adjusting encodings, or modifying chart types, while receiving contextual recommendations that encourage a deeper understanding of the dataset.

The system should act as a cognitive partner, identifying misconceptions and prompting users to assess their findings critically. For example, it could flag potential misinterpretations by explaining statistical limitations or suggesting alternative visualizations when an approach overlooks significant relationships. The system should integrate uncertainty visualizations or annotations that explain why certain conclusions may be misleading. 

Crucially, the AI must position itself as a coach rather than an omniscient authority. Instead of simply correcting its output, it should provide explanations that justify its suggestions and guide users toward independent decision-making. One approach could involve presenting multiple alternative visualizations alongside brief justifications for each, allowing users to compare perspectives and ask questions on the spot rather than passively accepting a single AI-generated output. Additionally, integrating a user-aware AI model could help tailor interactions to the user’s expertise level—for instance, providing more detailed explanations for novice users while offering concise, high-level prompts for experts.

To support the full lifecycle of visualization, the system should ensure that its output is aligned with the intended audience (and its predicted expertise) and medium. Whether the visualization is intended for a research poster, a local newspaper article, or an interactive dashboard for medical staff monitoring a patient's vitals, the AI could tailor design choices to suit the specific context and audience. For example, it could suggest adjustments to improve accessibility, recommend alternative visual encodings based on display constraints and legibility, or incorporate interactivity using the implementation language specified by the user.

Throughout this process, transparency remains a non-negotiable requirement. The system should not only justify its recommendations but empower users to control decisions, explore alternatives, and retain full creative and decision-making control. By balancing user-adaptive automation with human agency, we envision such an AI-driven visualization system evolving from a simple input-output tool into an interactive partner – one that adapts to the user's individual level, fosters critical thinking, deepens engagement and curiosity, and ultimately transforms how users interact with data.

\section{Reflection and Conclusion}
Our vision highlights the potential of expertise-aware adaptive systems in augmenting human cognition, demonstrating how GenAI-driven visualization tools could move beyond automation toward meaningful user engagement. By dynamically adapting to user expertise, such systems can promote deeper analytical reasoning and guide users through the visualization process in an interactive and transparent manner. However, our vision also highlights broader concepts that require further exploration, such as clearly defining and communicating the AI’s role and limitations. A truly human-centered system must balance bi-directional control: not only should the AI assist users in refining their analyses, but it should also empower and motivate users to critically challenge the AI’s outputs. This creates a dynamic exchange where both the user and AI contribute to a more meaningful engagement with data.

This position paper argues that expertise-aware systems, those that tailor their reasoning and guidance to the user, have the potential to enhance engagement with data. While we have focused on visualization, these considerations extend to other domains where GenAI supports sensemaking and decision-making. As AI continues to shape workflows across disciplines, ensuring that it enhances rather than replaces human cognition requires an interdisciplinary effort. Researchers in human-computer interaction, cognitive science, learning sciences, and AI must collaborate to design systems that not only support user needs but also encourage creativity, critical thinking, and reflective analysis.

%
%

\begin{acks}
Kathrin Schnizer was supported by the Deutsche Forschungsgemeinschaft (DFG, German Research Foundation), Project-ID 251654672-TRR 161 (C06).
\end{acks}

\bibliographystyle{ACM-Reference-Format}
\bibliography{bibliography}


%
%
\end{document}